\documentclass[smallabstract,smallcaptions]{dccpaper}

\usepackage{epsfig}
\usepackage{citesort}
\usepackage{amsmath}
\usepackage{amssymb}
\usepackage{color}
\usepackage{url}
\usepackage{multirow}
\usepackage[OT1]{fontenc}

\newlength{\figurewidth}
\newlength{\smallfigurewidth}

\begin{document}

\title
{\large
\textbf{Video-based compression for plenoptic point clouds}
}

\author{%
Li Li$^{\ast}$, Zhu Li$^{\ast}$, Shan Liu$^{\star}$, and Houqiang Li$^{\dag}$\\[0.5em]
{\small\begin{minipage}{\linewidth}\begin{center}
\begin{tabular}{ccc}
$^{\ast}$University of Missouri-KC & $^{\star}$Tencent America & $^{\dag}$USTC \\
5100 Rockhill Road & 661 Bryant St & No. 443 Huangshan Road \\
Kansas City, MO 64111, USA & Palo Alto, CA 94301, USA & Hefei, 230027, China\\
{lil1,lizhu}@umkc.edu & shanl@tencent.com & lihq@ustc.edu.cn
\end{tabular}
\end{center}\end{minipage}}
}

\maketitle
\thispagestyle{empty}

\begin{abstract}
The plenoptic point cloud that has multiple colors from various directions, is a more complete representation than the general point cloud that usually has only one color.
It is more realistic but also brings a larger volume of data that needs to be compressed efficiently.
The state-of-the-art method to compress the plenoptic point cloud is an extension of the region-based adaptive hierarchical transform (RAHT).
As far as we can see, in addition to RAHT, the video-based point cloud compression (V-PCC) is also an efficient point cloud compression method.
However, to the best of our knowledge, no works have used a video-based solution to compress the plenoptic point cloud yet.
In this paper, we first extend the V-PCC to support the plenoptic point cloud compression by generating multiple attribute videos.
Then based on the observation that these videos from multiple views have very high correlations, we propose encoding them using multiview high efficiency video coding.
We further propose a block-based padding method that unifies the unoccupied attribute pixels from different views to reduce their bit cost.
The proposed algorithms are implemented in the V-PCC reference software.
The experimental results show that the proposed algorithms can bring significant bitrate savings compared with the state-of-the-art method for plenoptic point cloud compression.
\end{abstract}

\section{Introduction}
\label{Sec::Introduction}
A point cloud is a set of 3D points that can be used to represent a 3D surface.
Each point is usually associated with one single color along with other attributes.
The point cloud can be used in many applications involving the rendering of 3D objects \cite{Tulvan2016} such as 3D immersive telepresence and 6 degree-of-freedom virtual reality.
However, the point cloud with only one single color is essentially not realistic since the colors of the real world objects may vary significantly along with the change of the view angles.
Recently, 8i captures several plenoptic point clouds with different colors in different view angles \cite{Tulvan2018}.
These point clouds are more realistic but also much larger, and thus bring more burdens to the communication and storage.
For example, the $12$ or $13$ attributes in the plenoptic point clouds make the attributes of the point clouds $12$ or $13$ times larger.
Therefore, there is an urgent need to compress them efficiently.

To transmit or store a point cloud, we need to signal the geometry or position information as well as attribute information.
Octree and its variations \cite{schnabel2006octree} are typical methods to compress the geometry.
Some methods also introduce plane \cite{kathariya2018scalable} or mapping \cite{He2017} to compress the geometry more efficiently.
However, as the main difference between the plenoptic point cloud and the general point cloud with one single color is the attribute information, we put more focuses on the review of the point cloud attribute compression in this paper.

The first group of works focusing on attribute compression is the transform-based method.
Zhang et al. \cite{Zhang2014} first proposed using Graph Fourier Transform (GFT)  to exploit the correlations among the already encoded geometry information.
However, deriving the transform kernel requires solving a very complex eigenproblem.
Therefore, Queiroz and Chou \cite{Queiroz2016} introduced the Region-based Adaptive Hierarchical Transform (RAHT) to obtain a better balance between the complexity and the performance.
This work is adopted in the geometry-based point cloud compression (G-PCC) standard \cite{Schwarz2018} and is the recommended algorithm to compress the static dense point cloud.
The second group of works focusing on attribute compression is the mapping-based method.
As a representative, Mammou et al. \cite{Schwarz2018} proposed a video-based point cloud compression (V-PCC) method to project the point cloud to 2-D videos patch by patch and compress them using high efficiency video coding (HEVC) \cite{Sullivan2012}.
Due to the high efficiency of the 2-D video compression standard, the coding efficiency of the V-PCC is very high.
It is shown in one recent work that the static dense point cloud can be also coded efficiently using V-PCC \cite{Gonçalves2019}.
The third group of works is the prediction-based method mainly designed for the sparse point cloud attribute.
Mammou et al. \cite{Schwarz2018} introduced a layer-based prediction to predict the point cloud from its coarse representation.
Kathariya et al. \cite{Kathariya2019} proposed using kd-tree to divide the point cloud into various layers to further improve the compression performance.

In terms of the compression of the plenoptic point cloud attributes, Sandri \emph{et al.} \cite{Sandri2019} extended the RAHT to further exploit the correlations among multiple attributes.
They used Discrete Cosine Transform (DCT) or Kahunen-Loeve Transform (KLT) to utilize the correlations among various attributes.
The RAHT-DCT or RAHT-KLT brings much better compression performance compared with the RAHT.
However, using only transform is unable to fully utilize the correlations among various views.
We believe a video-based solution with prediction, transform, quantization, and entropy coding is a better way to exploit the correlations.

Therefore, in this paper, we first extend the V-PCC to support the plenoptic point cloud by generating multiple attribute videos.
Under the current V-PCC framework, these attribute videos are encoded independently.
Then based on the observation that these videos from multiple views have very high correlations, we propose encoding them using multiview high efficiency video coding (MV-HEVC) \cite{Hannuksela2015} to utilize the correlations among various attributes.
Furthermore, as the unoccupied pixels will have no influences on the reconstructed quality of the plenoptic point cloud, we propose a block-based group padding method that unifies the unoccupied attribute pixels from different views to reduce their bit cost.

The rest of this paper is organized as follows.
In Section~\ref{Sec::plenoptic}, we will introduce the proposed video-based plenoptic point cloud compression framework.
The block-based padding for the unoccupied pixels will be introduced in Section~\ref{Sec::padding}.
Section~\ref{Sec::experiments} will describe the experimental results in detail.
Section~\ref{Sec::conclusion} will conclude the paper.

\section{Video-based plenoptic point cloud compression framework}
\label{Sec::plenoptic}
Under the V-PCC, a point cloud is first divided into several patches by projecting to its bounding box.
Each patch is generated by clustering the neighboring points with similar normals together.
In this way, the generated patches will have fewer variances in the geometry and attributes, and thus can be coded efficiently.
After the patches are generated, the V-PCC uses a simple packing strategy to organize the patches into frames.
The patch location is determined through an exhaustive search in a raster scan order.
In addition, patch rotation is supported to allow more flexible packing to improve compression performance.
After packing, the padding process aims to fill any empty space between the patches to make the generated frames more suitable for video coding.
The geometry and attribute videos will finally go through HEVC to generate the bitstream.
Note that each static point cloud is projected to two frames to handle the occlusion so as to achieve a better balance between the number of projected points and the compression performance.

To use the V-PCC to compress the plenoptic point cloud, we follow the above processes to generate the geometry and attribute videos.
The main difference is that there will be multiple attribute videos generated since each point has multiple attributes.
The multiple attribute videos will then be encoded using HEVC to be compressed efficiently.
Under this method, we will show the performance of the video-based plenoptic point cloud compression framework without using the correlations among various views.

\begin{figure}[t]
\begin{center}
\epsfig{width=3.5in,file=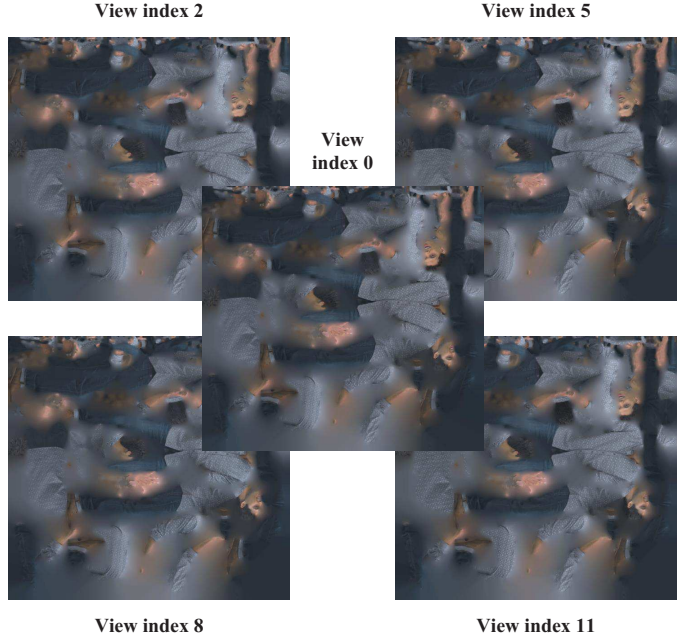}
\end{center}
\caption{\label{Fig::View}%
Some examples of the projected views from the plenoptic point cloud ``Loot''. These views are from the view index 0, 2, 5, 8, and 11, respectively.}
\end{figure}

\begin{figure}[t]
\begin{center}
\epsfig{width=4.5in,file=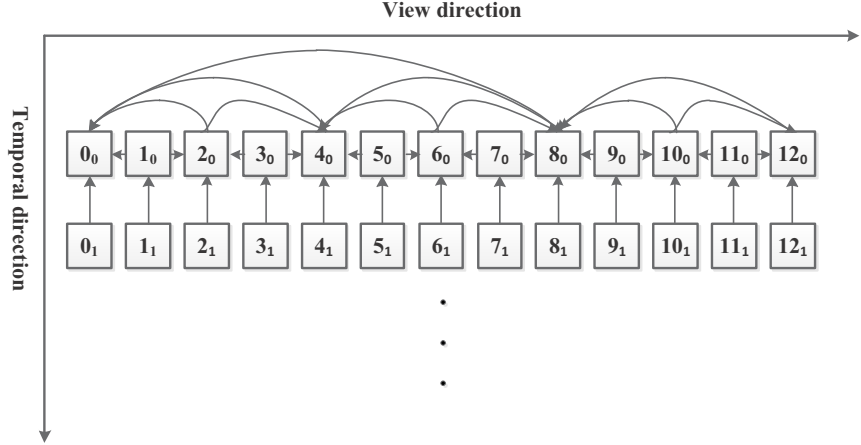}
\end{center}
\caption{\label{Fig::MV-HEVC}%
Proposed MV-HEVC-based plenoptic point cloud compression framework.}
\end{figure}

Fig.~\ref{Fig::View} gives a typical example of the attribute frames generated from multiple view angles.
We can see that the generated attribute frames from various view angles are very similar despite some pixel differences.
Therefore, we propose using MV-HEVC to exploit the correlations from multiple views to compress them more efficiently.
As we have mentioned, each static point cloud can be projected to two frames.
Therefore, using a $13$-view static point cloud as an example, we utilize the multiview encoding structure as shown in Fig.~\ref{Fig::MV-HEVC}.

In Fig.~\ref{Fig::MV-HEVC}, the indices from $0$ to $12$ indicate the index of each view.
The sub-indices $0$ and $1$ are the two frames projected from one point cloud.
The arrows among various squares in the figure indicate the reference relationships among various views and frames.
As shown in Fig.~\ref{Fig::MV-HEVC}, we organize a hierarchical coding structure with group of pictures (GOP) size $8$ in the view direction.
In this way, all the views are divided into $4$ hierarchical levels.
All the views are coded as B frames using bi-directional prediction to exploit the correlations among various views.
In the temporal domain, as the two frames projected from one static point cloud has much larger correlations than that among various views, the second frame will only reference the first frame with the same view index under the proposed reference structure.

\begin{table}[tp]
\begin{center}
\caption{\label{tab::QP-setting}%
QP settings of various views and frames}
{
\begin{tabular}{c|c|c}
\hline
hierarchical level & frame $0$ & frame $1$    \\
\hline
0 & $QP_I$+1 & $QP_I$+4 \\
1 & $QP_I$+2 & $QP_I$+5 \\
2 & $QP_I$+3 & $QP_I$+6 \\
3 & $QP_I$+4 & $QP_I$+7 \\
\hline
\end{tabular}}
\end{center}
\end{table}

In addition to the reference relationship, the quantization parameters (QPs) of various views and frames are also important to the compression performance.
Currently, we set the QPs according to the following basic rule.
First, the higher the hierarchical level is, the larger the QP is.
In the view direction, we set the QPs of each level as $QP_I + level + 1$.
Second, the second non-reference frame uses a higher QP than the first reference frame.
In the temporal direction, we set the QP of $QP_1$ as $QP_0 + 3$.
The detailed setting of the QPs of all the frames and views are shown in Table~\ref{tab::QP-setting}.
In this work, the proposed coding structure is used to compress static plenoptic point clouds.
This coding structure can be easily applied to dynamic plenoptic point clouds by propagating the coding structure in the temporal domain.

\section{Block-based group padding for unoccupied pixels}
\label{Sec::padding}
In the V-PCC, the unoccupied pixels are padded in a way that they will cost bits as few as possible since they have no influences on the reconstructed point cloud quality.
Similarly, in the plenoptic point cloud, we also need to minimize the bit cost of the unoccupied pixels.
In the V-PCC, several padding methods \cite{Jungsun2018} \cite{Sungryeul2018} are proposed to minimize the bit cost across the temporal direction.
Especially, all the unoccupied pixels are padded using the average of the first frame and the second frame to minimize the bit cost of the unoccupied pixels in the second frame.
However, since the pixel differences between the pixels in various views are much larger than that in the temporal domain, this method is unable to deal with the unoccupied pixels in the view directions.

\begin{figure}[t]
\begin{center}
\epsfig{width=2.5in,file=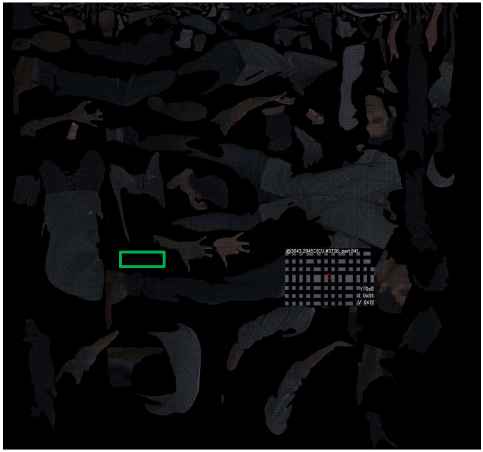}
\end{center}
\caption{\label{Fig::unoccupied}%
Typical example of a projected view with the unoccupied pixels set as black from the plenoptic point cloud ``Loot''.}
\end{figure}

Fig.~\ref{Fig::unoccupied} shows a typical example of the projected attribute with the unoccupied pixels set as black.
We can see that some large continuous unoccupied pixels exist as indicated by the green rectangle.
Those pixels can be padded properly to increase the correlations across various views to save some bits.
However, there are also some isolated unoccupied pixels as indicated by the small red square.
If we pad those pixels across the view direction, the blocks including both occupied and unoccupied pixels will become less continuous in the spatial domain.
This will make those blocks unable to find the corresponding block in the reference frame and lead to a serious bitrate increase.

Therefore, in this paper, we propose a block-based group padding for unoccupied pixels.
For each pixel, we first find a block with the pixel as the center pixel.
Only when all the pixels in the block are unoccupied, we will pad the pixel.
In this way, the isolated unoccupied pixels will not be padded and we can still keep the spatial continuity for the blocks including both occupied and unoccupied pixels.
For the large continuous unoccupied pixels, we can reduce the prediction residue so as to compress them more efficiently.
In our current implementation, the block size is set to $4$.

After the detection of the continuous unoccupied pixels, we will pad them using the average value of all the views in both frame $0$ and frame $1$,
\begin{equation}
f_{i,j} = \sum\limits_{k=0}^{N-1} ( f_{0,k} + f_{1,k}) / (2N), i \in {0,1}, j \in {0,1,...,N-1},
\end{equation}
where $N$ is number of views for the plenoptic point cloud, $i$ is the frame index, $j$ is the view index, and $f_{i,j}$ is the value to be padded for each position.
After the above padding scheme, in both the view direction and temporal direction, we can obtain a very good prediction for the continuous unoccupied pixels, and thus the bitrate of the unoccupied pixels can be significantly reduced.

\begin{table}[tp]
\begin{center}
\caption{\label{tab::character}%
Characteristics of the plenoptic point clouds}
{
\begin{tabular}{c|c|c|c|c}
\hline
Name & Points & Cameras &  Geometry bit depth & Attribute bit depth    \\
\hline
Boxer        &  3496011   &   13     &   12    &     8               \\
Loot         &  3021497   &   13     &   12    &     8               \\
Soldier      &  4007891   &   13     &   12    &     8               \\
Thaidancer   &  3130215   &   13     &   12    &     8               \\
Longdress    &  3100469   &   12     &   12    &     8               \\
Redandblack  &  2776067   &   12     &   12    &     8               \\
\hline
\end{tabular}}
\end{center}
\end{table}

\begin{table}[!htbp]
\begin{center}
\caption{\label{tab::state-of-the-art}%
Comparison between the proposed multi-view solution and RAHT-KLT \cite{Sandri2019}}
{
\begin{tabular}{c|cc|cc|c}
\hline
\multirow{2}{*}{Name}         &     \multicolumn{2}{c|}{RAHT-KLT}                &     \multicolumn{2}{c|}{Multiview-video-solution}              &    Luma     \\
                              &  Color bits &  Luma PSNR    &    Color bits   &   Luma PSNR             &   BD-rate   \\
\hline
\multirow{5}{*}{Box}          &   534974    &    36.58      &    594616     &      37.31               &    \multirow{5}{*}{--17.9\%}        \\
                              &   1102667   &    38.51      &    1142904    &      39.32               &              \\
                              &   2506516   &    41.02      &    2138928    &      41.37               &              \\
                              &   4144398   &    42.77      &    4185776    &      43.38               &              \\
                              &   7624336   &    45.08      &    8265288    &      45.24               &              \\
\hline
\multirow{5}{*}{Loot}         &   505156    &    36.47      &    521152     &      37.64               &    \multirow{5}{*}{--42.4\%}          \\
                              &   1036214   &    38.57      &    1005832    &      40.32               &              \\
                              &   2252251   &    41.16      &    1830264    &      42.71               &              \\
                              &   3576056   &    42.91      &    3332400    &      44.80               &              \\
                              &   6210303   &    45.21      &    6014088    &      46.54               &              \\
\hline
\multirow{5}{*}{Soldier}      &   1193244   &    34.15      &    966392     &      35.33               &    \multirow{5}{*}{--33.9\%}          \\
                              &   2361547   &    36.60      &    1867736    &      37.85               &              \\
                              &   3514995   &    38.24      &    3410864    &      40.14               &              \\
                              &   7227865   &    41.62      &    6146560    &      42.11               &              \\
                              &   11973133  &    44.15      &    11044128   &      43.82               &              \\
\hline
\multirow{5}{*}{Thaidancer}   &   434126    &    28.46      &    435816     &      31.26               &    \multirow{5}{*}{--50.4\%}           \\
                              &   1719585   &    33.63      &    823680     &      34.07               &              \\
                              &   3058823   &    36.63      &    1515208    &      36.66               &              \\
                              &   4292715   &    38.52      &    2842568    &      38.87               &              \\
                              &   5599587   &    40.03      &    5450840    &      40.91               &              \\
\hline
\multirow{5}{*}{Longdress}    &   519371    &    28.01      &    870840     &      32.98               &    \multirow{5}{*}{--40.6\%}           \\
                              &   2081546   &    33.01      &    1517448    &      35.36               &              \\
                              &   3770193   &    36.19      &    2575000    &      37.38               &              \\
                              &   5245716   &    38.36      &    4538608    &      39.15               &              \\
                              &   9214122   &    42.67      &    8311904    &      41.20               &              \\
\hline
\multirow{5}{*}{Redandblack}  &   224020    &    31.82      &    625080     &      36.48               &    \multirow{5}{*}{--28.3\%}           \\
                              &   903125    &    35.90      &    1098712    &      38.71               &              \\
                              &   1736193   &    38.43      &    1886296    &      40.69               &              \\
                              &   3313844   &    41.59      &    3416560    &      42.43               &              \\
                              &   6081458   &    45.08      &    6384704    &      44.08               &              \\
\hline
Average                       &    --       &     --        &     --        &      --                  &       --37.0\%                          \\
\hline
\end{tabular}}
\end{center}
\end{table}

\section{Experimental results}
\label{Sec::experiments}

The proposed algorithms are implemented in the V-PCC reference software TMC-7.0 \cite{TMC-7.0} to compare with the state-of-the-art method RAHT-KLT \cite{Sandri2019}.
We test all the static plenoptic point clouds defined in \cite{Tulvan2018}.
The characteristics of all the tested static plenoptic point clouds are shown in Table~\ref{tab::character}.
When verifying the performance of the proposed extension of V-PCC, we follow the V-PCC common test condition \cite{3DG} to test various bitrates from r1 (low bitrate) to r5 (high bitrate).
We use the all intra configuration to generate the results since we have only one static plenoptic point cloud to be compressed.
When verifying the performance of the proposed multiview-based solution, we use the same QP settings for the I frames as the V-PCC-based method.
The QPs of the other views and frames are set according to Table~\ref{tab::QP-setting}.
Note that the average peak signal to noise ratio (PSNR) of all the views is used as the quality metric for the attribute.

We first give a performance comparison between our multiview solution with the proposed padding method and the state-of-the-art method RAHT-KLT as shown in Table~\ref{tab::state-of-the-art}.
We can see that the proposed multiview solution can lead to $37.0\%$ performance improvements on average compared with the state-of-the-art method.
Through better utilizing the correlations among various views, the proposed method leads to a better compression performance compared with RAHT-KLT.
The experimental results demonstrate the effectiveness of the proposed algorithm.
In addition, we can see that the proposed algorithm always leads to a better R-D performance in low bitrate case.
While in high bitrate case, the proposed algorithm brings similar or even slightly worse R-D performance.

\begin{table}[tp]
\begin{center}
\caption{\label{tab::multiview_vs_VPCC}%
Comparison between the proposed multi-view solution and the V-PCC solution without using the correlations among multiple views}
{
\begin{tabular}{c|ccc|ccccc}
\hline
Name &  \multicolumn{3}{c|}{BD-AttrRate} & \multicolumn{5}{c}{BD-TotalRate}                    \\
                      &   \multicolumn{1}{c}{Luma}   &    \multicolumn{1}{c}{Cb}     &    \multicolumn{1}{c|}{Cr}     &   \multicolumn{1}{c}{D1}   &   \multicolumn{1}{c}{D2}    &    \multicolumn{1}{c}{Luma}    &   \multicolumn{1}{c}{Cb}  &   \multicolumn{1}{c}{Cr}       \\
\hline
Boxer        &     --62.4\%   &   --67.1\%    &   --69.2\%   &    --22.9\%  &  --22.6\%    &     --23.1\%      &   --26.7\%     &    --26.0\%   \\
Loot         &     --67.1\%   &   --71.8\%    &   --73.3\%   &    --24.5\%  &  --24.3\%    &     --22.7\%      &   --27.5\%     &    --27.4\%   \\
Soldier      &     --73.6\%   &   --75.1\%    &   --76.1\%   &    --33.3\%  &  --33.0\%    &     --32.6\%      &   --37.0\%     &    --37.2\%   \\
Thaidancer   &     --82.6\%   &   --83.5\%    &   --83.2\%   &    --78.5\%  &  --78.6\%    &     --77.3\%      &   --79.1\%     &    --78.8\%   \\
Longdress    &     --86.5\%   &   --86.6\%    &   --86.5\%   &    --51.2\%  &  --50.9\%    &     --53.0\%      &   --55.6\%     &    --55.4\%   \\
Redandblack  &     --78.1\%   &   --78.1\%    &   --79.1\%   &    --36.3\%  &  --36.0\%    &     --36.1\%      &   --41.1\%     &    --38.9\%   \\
\hline
Average      &     --74.4\%   &   --76.8\%    &   --77.7\%   &    --42.1\%  &  --41.9\%    &     --41.7\%      &    --45.2\%    &    --45.0\%    \\
\hline
\end{tabular}}
\end{center}
\end{table}

After an overall comparison with the state-of-the-art method, we then verify the performance of the proposed methods one by one.
Table~\ref{tab::multiview_vs_VPCC} gives a comparison between the proposed multi-view solution compared with the V-PCC solution without using the correlations among multiple views.
We can see that through utilizing the correlations among various views, we can achieve $74.4\%$, $76.8\%$, and $77.7\%$ bitrate savings on average for the Luma, Cb, and Cr components with respect to the attribute bits, respectively.
Although the proposed algorithm mainly targets the attribute compression, the proposed algorithm can reduce the total bits so as to bring performance improvements for both the geometry and attribute with respect to the total bits.
We can achieve $42.1\%$ and $41.9\%$ bitrate savings on average for the geometry under D1 and D2 measurements.
In addition, we can achieve an average of $41.7\%$, $45.2\%$, and $45.0\%$ performance improvements for the Luma, Cb, and Cr components, respectively.
The experimental results demonstrate that the proposed multiview compression framework can substantially exploit the correlations among various views so as to improve the performance.

\begin{table}[tp]
\begin{center}
\caption{\label{tab::padding}%
Performance comparison between the proposed multi-view solution with and without the block-based padding method}
{
\begin{tabular}{c|ccc|ccccc}
\hline
Name &  \multicolumn{3}{c|}{BD-AttrRate} & \multicolumn{5}{c}{BD-TotalRate}                    \\
                      &   \multicolumn{1}{c}{Luma}   &    \multicolumn{1}{c}{Cb}     &    \multicolumn{1}{c|}{Cr}     &   \multicolumn{1}{c}{D1}   &   \multicolumn{1}{c}{D2}    &    \multicolumn{1}{c}{Luma}    &   \multicolumn{1}{c}{Cb}  &   \multicolumn{1}{c}{Cr}       \\
\hline
Boxer        &     --18.7\%   &   --13.8\%    &   --16.5\%   &   --3.1\%   &  --3.0\%    &     --3.2\%      &   --2.5\%     &    --2.6\%   \\
Loot         &     --16.5\%   &   --15.7\%    &   --15.0\%   &   --2.5\%   &  --2.4\%    &     --2.6\%      &   --2.3\%     &    --2.1\%   \\
Soldier      &     --9.6\%    &   --7.7\%     &   --7.4\%    &   --1.6\%   &  --1.5\%    &     --1.8\%      &   --1.6\%     &    --1.4\%   \\
Thaidancer   &     --13.3\%   &   --12.2\%    &   --12.6\%   &    --9.6\%  &  --9.6\%    &     --9.6\%      &   --9.2\%     &    --9.5\%   \\
Longdress    &     --8.1\%    &   --8.2\%     &   --8.2\%    &    --1.4\%  &  --1.4\%    &     --1.7\%      &   --1.8\%     &    --1.8\%   \\
Redandblack  &     --13.6\%   &   --13.6\%    &   --14.1\%   &    --2.3\%  &  --2.3\%    &     --2.3\%      &   --2.7\%     &    --2.5\%   \\
\hline
Average      &     --13.3\%   &    --11.5\%   &   --13.6\%   &    --3.6\%  &  --3.6\%    &     --3.8\%      &    --3.5\%    &    --3.5\%    \\
\hline
\end{tabular}}
\end{center}
\end{table}

Table~\ref{tab::padding} shows the performance of the proposed block-padding method under our proposed multiview-based solution.
We can see that the proposed block-based padding can provide an average of $13.3\%$, $11.5\%$, and $13.6\%$ performance improvements with respect to only the attribute bits for the Luma, Cb, and Cr components, respectively.
With respect to the total bitrate, the proposed algorithm can bring over $3.6\%$ bitrate savings on average for both the geometry and attribute.
The experimental results demonstrate that through making the unoccupied pixel with higher correlations, the block-based padding algorithm can bring obvious bitrate saving for the unoccupied pixels so as to improve the overall performance.

\begin{figure}[t]
\begin{center}
\begin{tabular}{cc}
\epsfig{width=2.5in,file=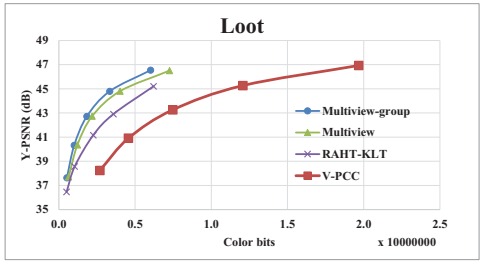} &
\epsfig{width=2.5in,file=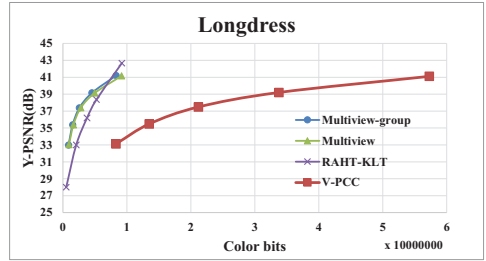} \\
{\small (a) Loot R-D curve} & {\small (b) Longdress R-D curve}
\end{tabular}
\end{center}
\caption{\label{Fig::RD_curve}%
Some examples of the R-D curves}
\end{figure}

Two examples of the R-D curves for all the above-mentioned methods are shown in Fig.~\ref{Fig::RD_curve}.
We can see that the V-PCC without considering the view correlations shows the worst performance.
The multiview-video solution shows a better performance than the RAHT-KLT, especially in the low bitrate case.
The proposed group padding algorithm can further improve the performance by saving the bit cost of the unoccupied pixels.

\section{Conclusion and future work}
\label{Sec::conclusion}
In this paper, we propose video-based plenoptic point cloud compression methods to compress the plenoptic point cloud more efficiently.
We first extend the current video-based point cloud compression (V-PCC) to support the plenoptic point cloud by generating multiple attribute videos.
Then based on the observation that these videos from multiple views have very high correlations, we propose encoding them using multiview high efficiency video coding.
We further propose a block-based group padding method that unifies the unoccupied attribute pixels from different views to reduce their bit cost.
The proposed algorithms are implemented in the V-PCC reference software.
The experimental results show that the proposed algorithms can bring significant bitrate savings compared with the state-of-the-art method for plenoptic point cloud compression.

\end{document}